\documentclass[useAMS,usenatbib,12pt]{mn2e}
\usepackage[dvipdf]{graphicx,color}
\usepackage[latin1]{inputenc}
\usepackage{graphics}
\usepackage{amsfonts}
\usepackage{amsmath}
\usepackage{multicol}
\usepackage{layout}
\usepackage{amssymb}
\usepackage{epsfig}
\usepackage[a4paper]{hyperref}
\DeclareGraphicsExtensions{.eps}

\title[An improved model for the formation times of dark matter haloes]
{An improved model for the formation times of dark matter haloes}
\author[C. Giocoli, J. Moreno, R. K. Sheth \& G. Tormen]
{Carlo Giocoli$^{1,2}$, Jorge Moreno$^{1}$,
Ravi K. Sheth$^{1}$ \& Giuseppe Tormen$^{2}$
\thanks{Email:
 \href{mailto:carlogiocoli@unipd.it}{carlo.giocoli@unipd.it},
 \href{mailto:jmoreno@physics.upenn.edu}{jmoreno@physics.upenn.edu},
 \href{mailto:shethrk@physics.upenn.edu}{shethrk@physics.upenn.edu},
 \href{mailto:giuseppe.tormen@unipd.it}{giuseppe.tormen@unipd.it}.
}\\
\\
$^{1}$University of Pennsylvania, 209 South $33^{rd}$ Street,
      Philadelphia, PA $19104-6396$, USA\\
$^{2}$Dipartimento di Astronomia, Università degli Studi di Padova,
      Vicolo dell'osservatorio 2 I-35122 Padova, Italy
}
\begin{document}
\date{}
\pagerange{\pageref{firstpage}--\pageref{lastpage}} \pubyear{2006}
\maketitle
\label{firstpage}
\begin{abstract}
A dark matter halo is said to have formed when at least half
its mass hass been assembled into a single progenitor.
With this definition, it is possible to derive a simple but
useful analytic estimate of the distribution of halo formation
times. The standard estimate of this distribution depends on
the shape of the conditional mass function---the distribution
of progenitor masses of a halo as a function of time.
If the spherical collapse model is used to estimate the progenitor
mass function, then the formation times one infers systematically
underestimate those seen in numerical simulations of hierarchical
gravitational clustering.
We provide estimates of halo formation which may be related to
an ellipsoidal collapse model. These estimates provide a
substantially better description of the simulations.
We also provide an alternative derivation of the formation time
distribution which is based on the assumption that haloes
increase their mass through binary mergers only.
Our results are useful for models which relate halo structure
to halo formation.
\end{abstract}


\section{Introduction}
Large samples of galaxy clusters selected in the optical
\citep{milleretal,mckayetal} and other bands (ACT, SPT)
will soon be available. These cluster catalogs will be used to
constrain cosmological parameters. The tightness of such constraints
depends on the accuracy with which the masses of the clusters can
be determined from observed properties.  These properties are
expected to depend on the formation histories of the clusters.
Models of cluster formation identify clusters with massive
dark matter halos, so understanding cluster formation requires
an understanding of dark halo formation.

There is also some interest in using the distribution of galaxy
velocity dispersions \citep{shethetal} to constrain
cosmological parameters \citep{nd02}.
The velocity dispersion of a galaxy is expected to be related to
the concentration of the halo which surrounds it, and this
concentration is expected to be influenced by the formation history
of the halo \citep{t98,betal01,wetal02}.
Hence, this program also benefits from understanding the formation
histories of dark matter halos.

In hierarchical models, the formation histories of dark matter halos
are expected to depend strongly on halo mass---massive halos are
expected to have formed more recently \citep{ps74}.
But quantifying this tendency requires a more precise definition of
what one means by the `formation time'.
\citet{lc93} provided a simple definition---it is the
earliest time when a single progenitor halo contains half the final
mass.  For this definition, they showed how to estimate the distribution
of halo formation times. \citet{st04} provide associated
expressions for the joint distribution of formation time and the mass
at formation.  This estimate depends on the distribution of
progenitor masses at earlier times, and Lacey \& Cole used the
assumption that halos form from a spherical collapse to estimate
this conditional mass function.  However, a model based on ellipsoidal
collapse provides a more accurate description of the abundances of
dark halos \citep{smt01} and of their progenitors
\citep{st02}.  Hence, one expects to find that the
ellipsoidal collapse model also provides a better description of
halo formation.  The primary goal of the present work is to study
if this is indeed the case.

Section~\ref{theory} reviews Lacey \& Cole's argument, and describes
how we estimate the ellipsoidal collapse based progenitor mass
functions.  Explicit formulae are provided in Appendix~\ref{messy}.
Section~\ref{sims} compares the model predictions for halo
formation with measurements in simulations of hierarchical
gravitational clustering.  The simulations, and how we estimate
halo formation times in them, are described in Section~\ref{gif2}.
Sections~\ref{pz} and~\ref{medianz} show that the spherical collapse
model predicts lower formation redshifts than are seen in the
simulations, particularly at low masses.  Insertion of ellipsoidal
collapse based expressions in Lacey \& Cole's formalism provides
substantially increased accuracy.
However, the agreement is not perfect---although the ellipsoidal
collapse based expressions provide a good description of how the
median formation redshift decreases with halo mass, it predicts
formation time distributions which are broader than seen in the
simulations.  A final section summarizes
our results and discusses possible reasons for the discrepancy.

There is some confusion in the literature about the relation
between halo formation as defined by \citet{lc93}, and
a quantity which arises in binary merger models of hierarchical
clustering.  The relation between this other quantity and the one
defined by Lacey \& Cole is clarified in Appendix~\ref{smol}.

\section{The formation time distribution}\label{theory}
Following \citet{lc93}, we will define the formation time
of a halo as the earliest time that at least half of its mass has
been assembled into a single progenitor.

\subsection{Relation to the progenitor mass function}\label{pznm}
Consider an ensemble of
halos of mass $M$ at time $T$, and let $N(m,t|M,T)$ denote the
average number of progenitors of these halos that have mass $m<M$
at time $t<T$.  Since a halo can have at most one progenitor of
mass $m>M/2$, the fraction of halos which have a progenitor of
mass $m>M/2$ at time $t$ is
\begin{displaymath}
 \int_{M/2}^M {\rm d}m\,N(m,t|M,T).
\end{displaymath}
But, because they have a progenitor of mass $m>M/2$, these halos
are also the ones which formed at some $t_{\rm f}<t$.  Hence,
\begin{equation}
 \int_0^{t_{\rm f}} {\rm d}t\,p(t|M,T) =
  \int_{M/2}^M {\rm d}m\,N(m,t_{\rm f}|M,T),
 \label{lc}
\end{equation}
where $p(t|M,T)$ denotes the probability that a halo of mass
$M$ at $T$ formed at time $t$.  Differentiating with respect to
$t_{\rm f}$ yields
\begin{eqnarray}
 p(t_{\rm f}|M,T) &=& \frac{\rm d}{{\rm d}t_{\rm f}}
                \int_{M/2}^M {\rm d}m\, N(m,t_{\rm f}|M,T) \label{ptMT}\\
  &=& \int_{M/2}^M {\rm d}m\,
      \frac{{\rm d} N(m,t_{\rm f}|M,T)}{{\rm d}t_{\rm f}}.
 \label{ptMTsmol}
\end{eqnarray}
Evidently, the formation time distribution is closely related to
the distribution of progenitor masses and its evolution.
Different estimates of the progenitor mass function will
result in different formation time distributions.

In what follows, we will estimate the distribution of halo
formation times using equation~(\ref{ptMT}).  Thinking along the
lines of equation~(\ref{ptMTsmol}) instead shows how
$p(t_{\rm f}|M,T)$ can be related to quantities which arise
naturally in binary merger models of hierarchical clustering.
Appendix~\ref{smol} provides details.

\citet{lc93} used estimates for $N(m,t|M,T)$ which were derived
from a model in which halos form from a spherical collapse
\citep{gg72,ps74}.  In this case, they showed that the distribution
of formation times could be scaled to a universal form:
\begin{equation}
 p(\omega) = 2\omega\,{\rm erfc}(\omega/\sqrt{2})
\end{equation}
where
\begin{equation}
 \omega = \sqrt{q}\, {\delta_{\rm sc}(z_{\rm f})-\delta_{\rm sc}(z_0)\over
                                  \sqrt{S(M/2)-S(M)}}
\label{littleomega}
\end{equation}
and $q=1$.  Here $\delta_{\rm sc}(z)$ is the overdensity required
for spherical collapse at $z$, and $S(M)$ is the variance in the
linear fluctuation field when smoothed with a top-hat filter of
scale $R = (3M/4\pi\bar\rho)^{1/3}$, where $\bar\rho$ is the
comoving density of the background.
In essence, $\omega$ is simply a scaled time variable:  in an
Einstein de-Sitter background cosmology $\omega\propto (z_{\rm f}-z_0)$,
where the constant of proportionality depends on the final mass $M$.

Strictly speaking, this expression is valid for a white-noise
power spectrum ($S(m) \propto  m^{-1})$, but it has been found to
provide a reasonable aproximation for more general power spectra
as well.  For this distribution, the median value of $\omega$ is
$0.974$, and the associated cumulative distribution is
\begin{equation}
 P(>\omega) = \sqrt{\frac{2}{\pi}}\,\omega\, \mathrm{e}^{-\omega^{2}/2}
  + (1-\omega^{2})\,\mathrm{erfc} \Big{(} \frac{\omega}{\sqrt{2}} \Big{)}\,.
 \label{pwlc}
\end{equation}

However, recent work has shown that the spherical collapse model
predicts fewer massive halos and more intermediate mass halos than
are seen in simulations of hierarchical gravitational clustering
\citep{st99}.  Models in which halos form from an
ellipsoidal collapse may be more accurate \citep{smt01}.
These models also provide a better description of the progenitor
mass function \citep{st02}.
In what follows, we will show the effects of substituting the
ellipsoidal collapse based expressions for $N(m,t|M,T)$ in
equation~(\ref{lc}).
We will also show that simply taking values of $q$ smaller than
one will make equation~(\ref{pwlc}) a good fit to the formation
redshift associated to the ellipsoidal collapse.

\begin{figure}
 \centering
 \includegraphics[width=\hsize]{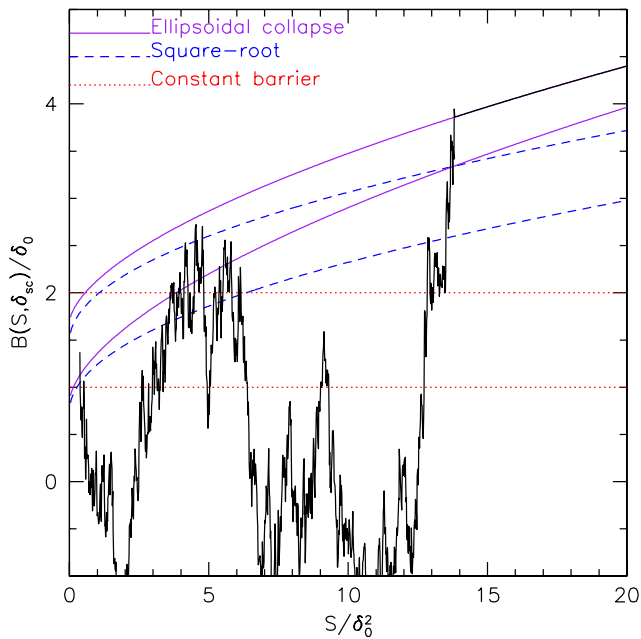}
 \caption{An example of a random walk and the three barrier
 shapes we consider in this paper.
 Here $\delta_{0}$ refers to the critical value associated 
 with spherical collapse overdenisity at redshift $z_{0}$.
 The jagged line is a sample brownian walk absorbed by the
 barrier associated with ellipsoidal collapse (solid curve).
 Short-dashed curves show the square-root barriers which we use
 to approximate the ellipsoidal collapse barrier, and horizontal
 dotted lines show the constant barrier associated with
 spherical collapse.  The upper set of barriers are associated
 with collapse at an earlier time.}
 \label{3barriers}
\end{figure}

\subsection{Progenitor mass functions in the excursion set approach}\label{ex}
We will use two approximations for the progenitor mass function.
Both forms are derived from casting the ellipsoidal collapse model
in the same language used for the spherical model---the excursion
set formalism of \citet{betal91}.
In this formalism, halo abundances at a given time are associated
with the first crossing distribution by Brownian motion random
walks, of a barrier whose height decreases with time, and may
in addition depend on how many steps the walk has taken.
Figure~\ref{3barriers} illustrates:  it shows a walk which starts
at some initial $S(M)$, and walks to the right, eventually crossing
the barrier at some larger value of $S(m)>S(M)$.
Here $S(m)$ is the variance in the initial fluctuation field when
smoothed with a tophat filter of comoving scale
 $R = (3m/4\pi\bar\rho)^{1/3}$,
($\bar\rho$ is the comoving background density),
and the height of the walk represents the value of the initial
density field when smoothed on scale $R$. 
In hierarchical models, $S(m)$ is a monotonically decreasing
function of mass $m$.
Hereafter we will use $\delta_{0}$ and $\delta_{1}$ to indicate the overdensity associated 
with the spherical collapse model respectively at redshift $z_{0}$ and $z_{1}$.

In the excursion set approach, the shape of the barrier, $B(S)$,
depends on the collapse model.  If the barrier is crossed on scale
$S(m)$, then this indicates that the mass element on which the sphere
is centered will be part of a collapsed object of mass $m$.
Bond et al. used the fraction of walks which
cross $B(S)$ at $S(m)$ as an estimate of the fraction of mass in
halos of mass $m$:  the parent halo mass function.
Similarly, the fraction $f(s|S)$ of walks which start from some
scale $S(M)$ and height $B(S)$ and first cross the barrier $B(s)$
at some $s>S$, can be used to provide an estimate of the
progenitor mass function of $M$ halos:
\begin{equation}
 N(m|M)\,{\rm d}m
   = \left(\frac{M}{m}\right)\,f(s|S)\,{\rm d}s
 \label{nmM}
\end{equation}
(note that $m<M$ because $s>S$).
Thus, in the excursion set approach, the shape of $N(m|M)$ depends
on how the shape of the barrier changes with mass---it is in this
way that the collapse model affects the parent halo and progenitor
mass functions.

The barrier shape is particularly simple for the spherical model:
$B=\delta_{\rm sc}$ is the same constant for all values of $S$.
\citet{smt01} argued that the barrier associated with ellipsoidal 
collapse scales approximately as
\begin{equation}
 B(S) = \sqrt{q}\delta_{\rm sc}
   \left\{1 + \beta \left[{S(m)\over q\delta_{\rm sc}^2}\right]^\gamma\right\},
 \label{smt}
\end{equation}
where $\delta_{\rm sc}$ is the value associated with complete
collapse in the spherical model.  Following Sheth et al., 
we set $\beta = 0.5$, $\gamma=0.6$ and $q=0.707$; the values of 
the first two parameters are motivated by an analysis of the collapse
of homogeneous ellipsoids, whereas the value of $q$ comes from
requiring that the predicted halo abundances match those seen in 
simulations.  
The standard spherical model has $\beta=0$ and $q=1$; we will also
show the predictions of this model if $\beta=0$ but $q=0.707$.

For general $\gamma$, exact analytic solutions to the first
crossing distribution of barriers given by equation~(\ref{smt})
are not available.  For this reason, we have studied two analytic
approximations which result from addressing the first crossing
problem in two different ways.
The first approach uses an analytic approximation to the first
crossing distribution which \citet{st02} showed was
reasonably accurate.

Our second approach is to substitute the ellipsoidal collapse
barrier for one which is similar, but for which an exact analytic
expression for the first crossing distribution is available.
Specifically, when $\gamma = 0.5$, then the barrier height
increases with the square-root of $S$, and the first crossing
distribution can be written as a sum of parabolic cylinder
functions (Breiman 1966).  In this approach, we approximate the
progenitor mass function using an expression which would be
exact for a square-root barrier, bearing in mind that the
square-root barrier is an approximation to the ellipsoidal
collapse dynamics.  For this barrier shape, we set
$\beta = 0.5$ and $q=0.55$ since these choices result in
parent halo and progenitor mass functions which best fit the
simulations; Figure~\ref{3barriers} illustrates the similarity
of the barrier shapes.
We will show that both these approximations provide substantially
better descriptions of the simulations than does the expression
which is based on spherical collapse.
Explicit expressions for these barrier crossing distributions
are provided in Appendix~\ref{messy}.  They are related to the
progenitor mass function by equation~(\ref{nmM}).

\section{Comparison with simulations}\label{sims}
The simulation data analysed below are publicly available at
\href{http://www.mpa-garching.mpg.de/Virgo}
{\texttt{http://www.mpa-garching.mpg.de/Virgo}}.
The simulation itself is known as GIF2, and is described in some
detail by \citet{getal04}.
It is of a flat $\mathrm{\Lambda}$CDM universe with parameters
($\Omega_m,\sigma_8,h,\Omega_bh^2) = (0.3,0.9,0.7,0.0196)$.
The initial fluctuation spectrum had an index $n=1$, with
transfer function produced by CMBFAST \citep{sz96}.

\begin{figure*}
 \centering
 \includegraphics[width=\hsize]{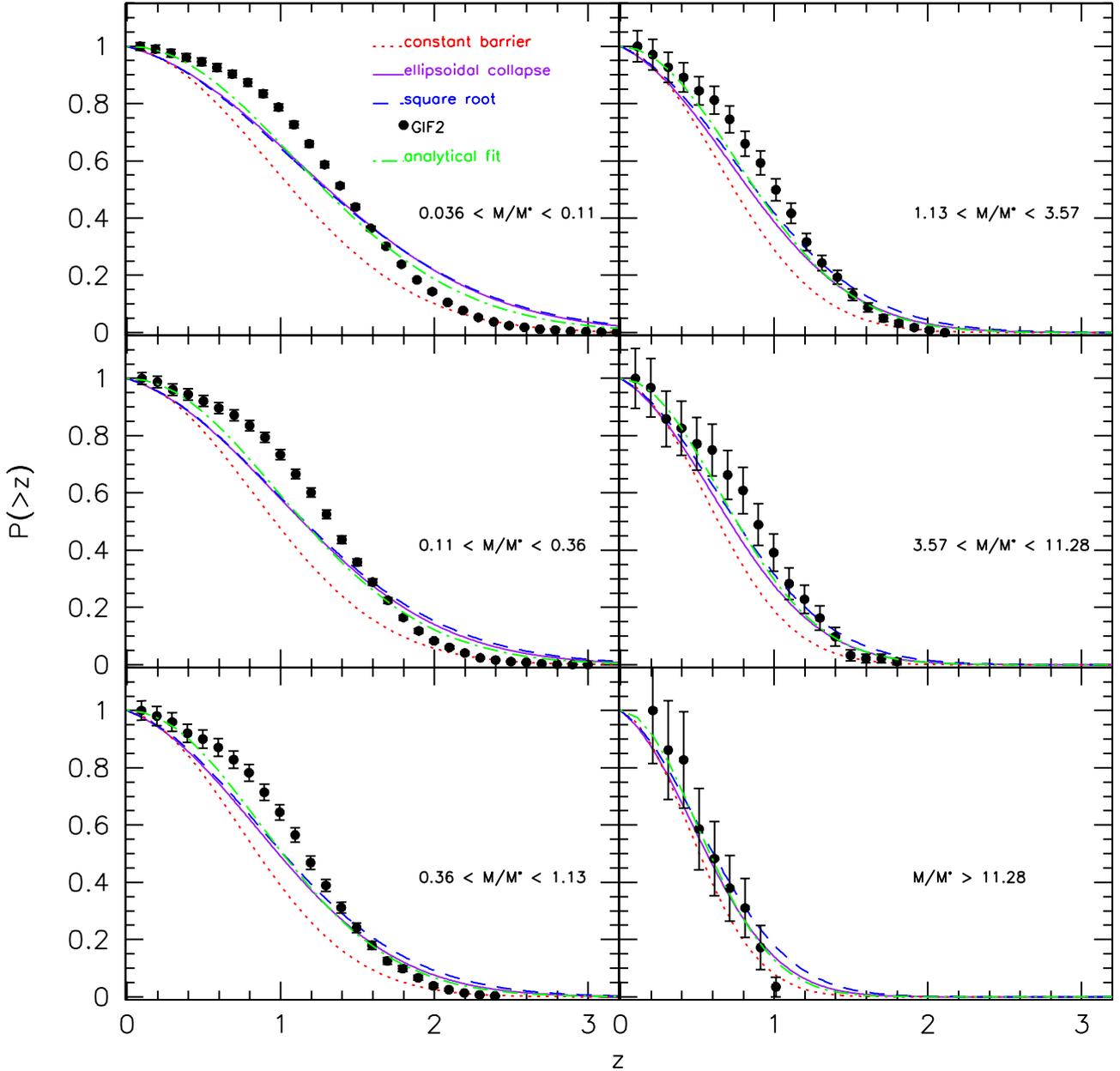}
 \caption{Cumulative distribution of dark halo formation times
          for halos identified at $z=0$.
          From top to bottom, panels show results for halos
          with masses in the range log$_{10}M/h^{-1}M_{\odot}:$
          11.5-12, 12-12.5, 12.5-13, 13-13.5, and 13.5-14.
          Symbols show the measurements in GIF2; dotted curve shows
          the prediction associated the constant barrier spherical collapse
          model; dot-dashed curve shows the analytical fit equation~(\ref{pwlc})
          with $q=0.707$.
          Short-dashed and solid curves show the predictions associated
          with the square-root and ellipsoidal collapse based models. }
 \label{cform0}
\end{figure*}

\subsection{The GIF2 simulation}\label{gif2}
The GIF2 simulation followed the evolution of $400^3$ particles
in a periodic cube 110~$h^{-1}$Mpc on a side.
The individual particle mass is $1.73 \times 10^{9}\,h^{-1}M_\odot$,
allowing us to study the formation histories of haloes which are
one order of magnitude smaller in mass than was previously possible.
Particle positions and velocities were stored at 50 output times,
logarithmically spaced between $1+z=20$ and $1+z=1$.
Halo and subhalo merger trees were constructed from these outputs
as described by \citet{tmy04}.  Briefly, halos at a given redshift
are identified as spherical objects which enclose the virial density
appropriate the cosmological model at that redshift (computed using
the spherical collapse model).

A merger history tree was constructed for each halo more massive
than $10^{11.5}\,h^{-1}M_\odot$ (i.e., containing more than 180
particles) as follows.
The progenitors of a halo of mass $M_0$ at $z_0$ are found by
identifying in the previous output $z_1 > z_0$ all haloes which
contribute any number of particles to $M_0$.
Of these, the progenitor which contributes the most mass to $M_0$
is called $M_1$.  This procedure is then repeated, starting with
$M_1$ at $z_1$, considering its progenitors in the previous output
time $z_2>z_1$, and choosing from them the one, $M_2$, which
contributes the most mass to $M_1$.  In this way, the mass of the
most massive progenitor is traced backwards in time, to high redshift.

\begin{figure*}
 \centering
 \includegraphics[width=\hsize]{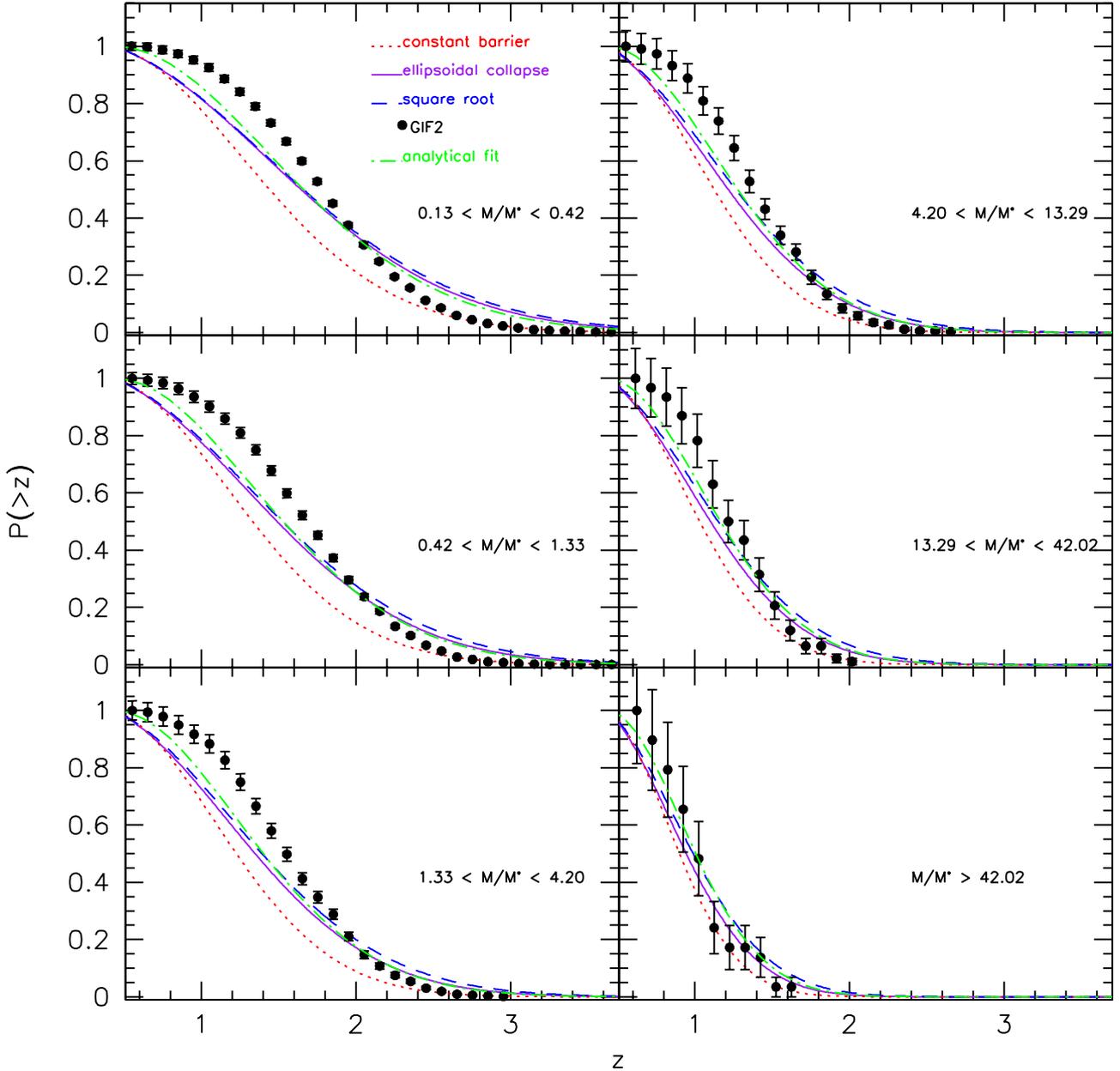}
 \caption{Same as previous figure, but now for halos identified
          at $z=0.5$.  As before, the spherical model predicts
          lower formation redshifts than observed.  Whereas
          the ellipsoidal collapse based expressions predict
          the median formation redshift reasonably well, they
          predict a broader range of redshifts than are observed.}
 \label{cform0.5}
\end{figure*}

We will present our results as a function of halo mass.
We have estimated formation times for
$5611$ halos with log$_{10} (M/h^{-1}M_\odot)$ in the range $11.5-12$,
$2431$ in the range $12-12.5$, $892$ in the range $12.5-13$,
$341$ in the range $13-13.5$, $92$ in the range $13.5-14$,
and $29$ in the range $14-14.5$.  These abundances are well fit
by the formula of \citet{st99}, so they are in better
agreement with the ellipsoidal rather than spherical collapse model.

\subsection{Cumulative distribution of formation times}\label{pz}

Figure~\ref{cform0} shows the cumulative distribution of dark
halo formation redshifts (i.e., equation~\ref{lc}) for halos
identified at $z=0$.
In general the analytical expressions obtained from the two 
different approximate solutions to the ellipsoidal collapse 
barrier problem---the \citep{st02} approximation for the first 
crossing distribution of the ellipsoidal collapse barrier, 
or the exact expression for the first crossing distribution of 
the square root barrier---are complicated. 
However, we have found that the predictions of the two ellipsoidal 
based models are quite well approximated by the expression for the 
spherical model with $n=0$, equation~(\ref{pwlc}), by simply changing 
the value of $q$.

Different panels show results for the mass bins described above.
The points show measurements in the GIF2 simulation, and the
four curves show the formation time distributions associated
with the $\mathrm{\Lambda}$CDM spherical collapse (dotted),
ellipsoidal collapse (solid), square-root barrier approximation
(dashed) and with equation~(\ref{pwlc}) using $q=0.707$ (dot-dashed).

Again, the spherical collapse model severely underestimates the
redshifts of halo formation.
The two ellipsoidal collapse based estimates fare better, in the
sense that they predict median formation redshifts which are
closer to those seen in the simulation.

However, both ellipsoidal based estimates clearly predict a wider
range of formation redshifts than is seen in the simulation---at
fixed mass, the distribution of halo formation redshifts is narrower
than predicted.  This remains true for the formation time distribution
of halos identified at $z=0.5$ shown in Figure \ref{cform0.5}.

\begin{figure}
 \centering
 \includegraphics[width=\hsize]{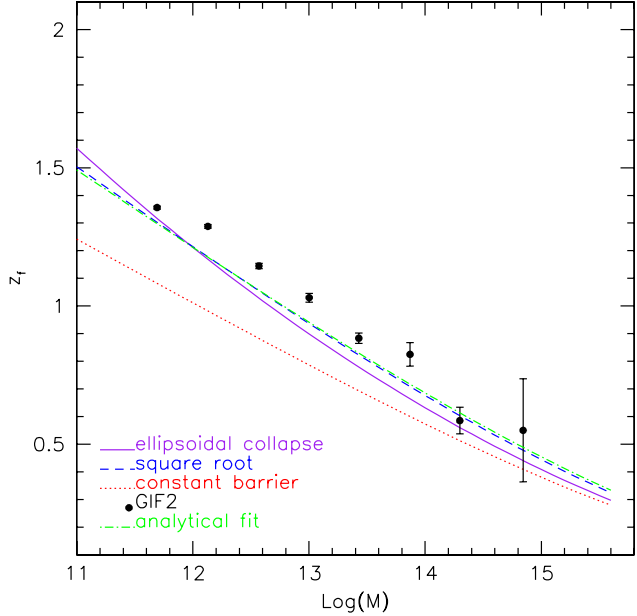}
 \caption{Median formation redshift for halos identified at $z=0$
          as a function of halo mass.  Points with (Poisson) error
          bars show our measurements in the GIF2 simulation.
          Smooth curves show the median formation times associated
          with three different models of halo formation:
          spherical collapse (dotted), ellipsoidal collapse (solid)
          and the square-root barrier approximation (short-dashed).
          Dot-dashed curve shows the prediction of equation~(\ref{mzfreleq})
          with $q=0.707$.}
 \label{medianzf}
\end{figure}

\subsection{Mass-dependence of median formation time}\label{medianz}
Figure~\ref{medianzf} shows the median formation redshift of
halos identified at $z=0$ as a function of halo mass.
Points show our measurements in the GIF2 simulation
The simulation showes that massive halos formed more recently:
the median formation redshift decreases with halo mass.

\begin{figure}
 \centering
 \includegraphics[width=\hsize]{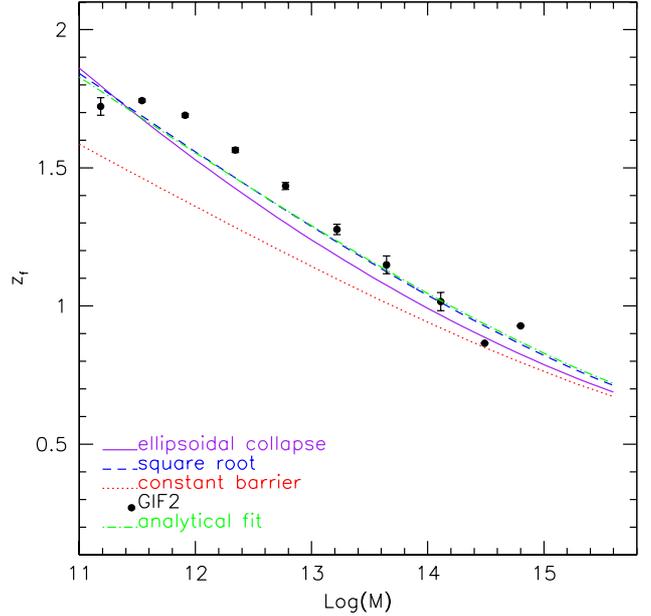}
 \caption{Same as previous Figure, but for halos
          identified at $z=0.5$ in the GIF2 simulation.}
  \label{medianz5f}
\end{figure}

Smooth curves show the formation time distributions associated
with three different models of halo formation: $\mathrm{\Lambda}$CDM
spherical collapse (dotted), ellipsoidal collapse (solid) and
the square-root approximation (short-dashed).
In all cases, the model predictions for the median formation
redshift were obtained by finding that $\bar{z}_f$ at which
\begin{equation}
 \int_{M/2}^M {\rm d}m\,N(m,z_{\rm f}|M,z) = {1\over 2}.
\end{equation}
The Figure shows that halos less massive than $10^{14}h^{-1}M_\odot$
clearly form at higher redshifts than predicted by the spherical
collapse model (dotted line).  Both our estimates of the ellipsoidal
collapse prediction are in substantially better agreement with the
simulations.
The prediction associated to equation~(\ref{pwlc}) is given implicitely
by inserting the median rescaled formation redshift $\bar{\omega} = 0.974$
into equation~(\ref{littleomega}), providing:
\begin{equation}
 \delta_{\rm sc}(\bar{z}_{\rm f}|M,z_0) =
 \delta_{\rm sc}(z_{0}) + {0.974\over \sqrt{q}}\, \sqrt{S(M/2) - S(M)}.
   \label{mzfreleq}
\end{equation}
The choice $q=0.707$ is shown by the dot-dashed line in the Figures;
setting $q=0.6$ yields even better agreement with the simulations.

Fitting functions, accurate to a few percent, for
$\delta_{\rm sc}(z)$ and $S(m)$ are available in the literature.
For instance,
\begin{equation}
 \delta_{\rm sc}(z) \approx
   D_+(z)\, \frac{3}{20}\, (12 \pi)^{2/3}\, \Omega_m^{0.0055}\,
\end{equation}
\citep{nfw97}, where $\Omega_m$ is the ratio of the background
to the critical density at $z$, and
\begin{equation}
D_{+}(z) = \frac{5}{2}\, \Omega_{m}
 \Big{[} \Omega_{m}^{4/7} - \Omega_{\mathrm{\Lambda}}
 + \Big{(}1+\frac{1}{2} \Omega_{m} \Big{)} \Big{(} 1 + \frac{1}{70}
 \Omega_{\mathrm{\Lambda}} \Big{)} \Big{]}^{-1}
\end{equation}
\citep{cpt92}, and, for a CDM power spectrum,
\begin{equation}
 S(m) = A \left(1 + 2.208 \bar{m}^{p}-0.7668 \bar{m}^{2p}
                  + 0.7949 \bar{m}^{3p}\right)^{-4/(9p)}
\end{equation}
\citep{TS00},
where $A$ is the normalization factor of the linear theory
power spectrum at $z=0$ (so it depends on $\sigma_{8}$),
$\bar{m} = m (h \Gamma)^{2}/10^{12} M_{\odot}$, where $\Gamma$
is the parameter which describes the shape of the power
spectrum (typically $\Gamma\approx \Omega h$), and $p=0.0873$.

\section{Discussion and Conclusions}
\citet{lc93} defined the formation time of an object
as the earliest time when at least half its mass was assembled
into a single progenitor.
We clarified the relation between this definition of formation
time, and a quantity which arises in binary merger models of
clustering (Appendix~\ref{smol}).
We have shown that insertion of spherical collapse based
expressions in Lacey \& Cole's (1993) formalism for halo
formation underestimates the redshifts of halo formation; 
this is consistent with previous work (Lin, Jing \& Lin 2003).
Ellipsoidal collapse based expressions are a marked improvement:
although they result in formation redshift distributions which
are broader than seen in simulations (Figures~\ref{cform0}
and~\ref{cform0.5}), they predict the median formation redshift
quite well (Figure~\ref{medianzf} and~\ref{medianz5f}).

The fact that our predicted formation time distributions are
broader than those seen in the simulation can be traced back to
the fact that the low redshift progenitor mass functions from the
excursion set approach are not in particularly good agreement with
the simulations.
This is similar to the findings of \citet{st02}, who
noted that some of the discrepancy was almost certainly due to
the idealization that the steps in the excursion set walks are
uncorrelated.   Recent work has shown that there is some
correlation between halo formation and environment
\citep{st04,getal05,hetal06,wetal06}---this almost certainly
indicates that a model with uncorrelated steps will be unable to
provide a better description of the simulations.
Nevertheless, the fact that the median formation redshifts
are quite well reproduced by our model does represent progress.

Although the exact expressions associated with our models are
complicated, we were able to find a useful fitting formula for the
median formation redshift, equation~(\ref{mzfreleq}), which we hope
will be useful in studies which relate the formation times of halos
to observable quantities.

\section*{acknowledgments}
We thank L. Marian, E. Ricciardelli, F. Biondi and E. Tescari for
helpful discussions, and the Aspen Center for Physics for
hospitality during the completion  of this work, which was supported
in part by NSF Grant 0520647 and by CONACyT--Mexico.

%
\bibliographystyle{mn2e}


\appendix
\section{Explicit expressions for barrier crossing distributions}\label{messy}
We are interested in walks which start from one barrier $B_0(S)$
and walk to another, $B_1(s)$, with $s>S$.  Thus, the barrier to
be crossed is of the form
\begin{equation}
 B_1(s) - B_{0}(S) = \sqrt{q}\delta_{1}
  \Big{[}1+\beta\Big{(}\frac{s}{q\delta_{1}^{2}}\Big{)}^{\gamma}\Big{]}
    - \sqrt{q}\delta_{0}
  \Big{[}1+\beta\Big{(}\frac{S}{q\delta_{0}^{2}}\Big{)}^{\gamma}\,\Big{]}.
\end{equation}
In the formulae $\delta_{0}$ and $\delta_{1}$ refer to the sferical
collapse overdenisity at redshift $z_{0}$ and $z_{1}$.
For barriers of this form, \citet{st02} showed that
\begin{eqnarray}
f(s|S)ds &\approx& \frac{\mathrm{d}s}{s-S}\,
                   \frac{|T(s|S)|}{\sqrt{2\pi (s-S)}} \nonumber\\
& &\times\quad
   \exp{\Bigg{\{} -\frac{[B_{1}(s)-B_{0}(S)]^{2}}{2(s-S)} \Bigg{\}} },
\label{ecst}
\end{eqnarray}
where
\begin{eqnarray}
 T(s|S)=\sum^{5}_{n=0}\frac{(S-s)^{n}}{n!}
  \frac{\partial^{n}[B_{1}(s)-B_{0}(S)]}{\partial s^{n}}\,.
\end{eqnarray}
When $\beta=0$ and $q=1$, this reduces to the spherical collapse
based expression used by \citet{lc93}:
\begin{eqnarray}
f(s|S)ds = \frac{\mathrm{d}s}{s-S}\,
           \frac{\delta_{1}-\delta_{0}}{\sqrt{2\pi (s-S)}}\,
  \exp\left\{-\frac{(\delta_{1}-\delta_{0})^{2}}{2(s-S)}\right\}.
\end{eqnarray}
When $\gamma=1/2$, then
\begin{equation}
 B_1(s) - B_{0}(S) = \sqrt{q}(\delta_1-\delta_0) - \beta\sqrt{S}
                      + \beta\sqrt{s},
\end{equation}
and the exact solution to this square-root barrier problem is
\begin{eqnarray}
f(s|S)ds &=& \exp \left({w_\beta^{2}-\beta^{2}\over 4}\right) \times
  \nonumber \\
 & &\sum_\lambda \Bigg{(}\frac{S}{s}\Bigg{)}^{\frac{\lambda}{2}}
  \frac{D_{\lambda}(w_{\beta})\,D'_{\lambda}(-\beta)}{I_{\lambda}(-\beta)}
  \frac{{\rm d}s}{2s}
 \label{sqrt}
\end{eqnarray}
\citep{breiman,mr05}, where
\begin{displaymath}
 w_{\beta} \equiv \frac{\sqrt{q}(\delta_{1}-\delta_{0})}{\sqrt{S}}-\beta,
 \quad
 I_\lambda(-\beta) \equiv
 \int^{\infty}_{-\beta} {\rm d}x\,D_\lambda^2(x),
\end{displaymath}
and $D_\lambda(x)$ are parabolic cylinder functions which
satisfy
\begin{displaymath}
 D'_\lambda(x) = (x/2)\, D_\lambda(x) - D_{\lambda+1}(x).
\end{displaymath}
The sum is over the roots $\lambda$ defined by
\begin{displaymath}
 D_\lambda(-\beta)=0, \quad {\rm so}\quad
 D'_{\lambda}(-\beta) = - D_{\lambda+1}(-\beta).
\end{displaymath}

\section{On the difference between halo formation and creation}\label{smol}
The main text deals with halo formation as defined by
\citet{lc93}.  Before continuing, it is worth mentioning
that there is another context in which the term `halo formation'
arises---this is when the halo population is modeled as arising
from a binary coagulation process of the type first described
by \citet{smol16}.
In this description, the time derivative of the halo mass
function is thought of as the difference of two terms:
one represents an increase in the number of halos of mass $m$
from the merger of two less massive objects, and the other is the
decrease in the number of $m$-haloes which results as $m$-haloes
themselves merge with other halos, creating more massive halos as
a result:
\begin{equation}
 \frac{{\rm d}n(m,t)}{{\rm d}t} = C(m,t) - D(m,t)
\end{equation}
The gain term, the first on the right hand side above,
is sometimes called the halo formation \citep{sp97}
or creation \citep{pm99,s03} term.
In what follows, we will use the word `creation' to mean this
term, and `formation' to mean the quantity studied by Lacey \& Cole.

Halo formation and creation are {\em very} different quantities,
as we show below.  Nevertheless, they are sometimes used
interchangeably in the literature (e.g. \citet{vkmb01}).
Here we show explicitly how to compute the Lacey-Cole formation
time distribution from the Smoluchowski-type creation and
destruction terms, with the primary aim of insuring that the error
of confusing one for the other is not repeated.

In the Lacey-Cole picture, the formation time distribution of
$M$-halos identified at $T$ is given by equation~(\ref{ptMT}).
In this case, one first integrates over the conditional mass
function, and then takes a time derivative.
Here, we will instead take the time derivative inside the
integral over mass and compute it before integrating over mass,
as suggested by equation~(\ref{ptMTsmol}).
In this case, the integrand has the form of a time derivative which
plays a central role in the Smoluchowski picture.  In particular,
we can write
\begin{equation}
 \frac{{\rm d}N(m,t|M,T)}{{\rm d}t} = C(m,t|M,T) - D(m,t|M,T),
\end{equation}
where $C$ and $D$ are now the creation and destruction terms
associated with the progenitor mass function, evaluated explicitly
in \citet{s03}.

This simple step shows explicitly that $p(t|M,T)$ can be written
in terms of Smoluchowski-like quantities:
\begin{equation}
 p(t|M,T) = \int_{M/2}^M {\rm d}m\, C(m,t|M,T)
            - \int_{M/2}^M {\rm d}m\, D(m,t|M,T).
\end{equation}
A little thought shows why this works out so easily.
Simply integrating the halo creation rate $C(m,t|M,T)$ over the
range $M/2\le m\le M$ overestimates halo formation, since some of
the halos created at $t$, with mass $3M/4\le m\le M$ say, may
actually have been may created by binary mergers in which one of
the halos had a mass in the range $M/2\le m\le 3M/4$.
Such creations should not be counted towards halo formation, since
formation refers to the earliest time that $m$ exceeds $M/2$.
However, it is precisely this double-counting which the second term,
the integral of $D(m,t|M,T)$ over the range $M/2\le m\le M$,
removes.  Note in particular that, whereas  the formation time
distribution $p(t|M,T)$ is related to Smoluchowski-type quantities,
it is {\em not} the same as $C(m,t)$.  For completeness, note that
\begin{equation}
 C(m,t) = \int_m^\infty {\rm d}M\,n(M,T)\,C(m,t|M,T),
\end{equation}
where $n(M,T)$ denotes the number density of halos of mass $M$ at
time $T$.

\label{lastpage}
\end{document}